\documentclass[prl,showpacs,twocolumn,amsmath,amssymb]{revtex4}
\usepackage{graphicx}% Include figure files
\usepackage{bm}% bold math

\def\br{\mathbf{r}}
\def\bk{\mathbf{k}}
\def\bq{\mathbf{q}}
\def\bQ{\mathbf{Q}}
\def\bR{\mathbf{R}}

\def\re{\mathrm{Re}\,}

\begin{document}

\title{Nonuniform mixed-parity superfluid state in Fermi gases}

\author{K. V. Samokhin$^{1}$ and M. S. Mar'enko$^{2}$ }

\affiliation{$^{1}$ Department of Physics, Brock University,
St.Catharines, Ontario, Canada L2S 3A1\\
$^{2}$Department of Physics and Astronomy, Hofstra University,
Hempstead, New York 11549, USA}
\date{\today}

\begin{abstract}
We study the effects of dipole interaction on the superfluidity in
a homogeneous Fermi gas with population imbalance. We show that
the Larkin-Ovchinnikov-Fulde-Ferrell phase is replaced by another
nonuniform superfluid phase, in which the order parameter has a
nonzero triplet component induced by the dipole interaction.
\end{abstract}


\maketitle

Bardeen-Cooper-Schrieffer (BCS) theory is the standard model of
superconductivity and superfluidity in fermionic systems. The
central concept of the BCS model is that of a Cooper pair composed
of fermions with opposite momenta and spins. If there is a
mismatch between the Fermi surfaces of spin-up and spin-down
fermions, e.g. because of the Zeeman splitting in magnetic field,
then the formation of the Cooper pairs costs energy, which results
in the suppression of the critical temperature. It was shown by
Larkin and Ovchinnikov \cite{LO64} and Fulde and Ferrell
\cite{FF64} (LOFF) that the pair-breaking effect of the Fermi
surface splitting can be reduced if the Cooper pairs have a
nonzero center-of-mass momentum. The resulting nonuniform
superconducting state turns out to be more favorable at low
temperatures than the uniform BCS state.

Due to the sensitivity of the nonuniform state to disorder and the
orbital effects, the attempts to find it in superconductors have
been unsuccessful (at least until the recent observation of the
features consistent with the LOFF state in the phase diagram of
the heavy-fermion compound CeCoIn$_5$ \cite{CeCoIn5}). The recent
surge of interest to the nonuniform superconducting and superfluid
states, see e.g. Refs. \cite{Comb01,MMI05,SR06,SMPM05,SM06,KJT06},
has been stimulated by the experimental progress in ultracold
atomic Fermi gases, such as ${}^6$Li and ${}^{40}$K. The
Fermi-surface splitting in these systems is due to the populations
of atoms in different hyperfine states being unequal
\cite{Zwier06,Part06}. When a pairing interaction between the
fermions is turned on, the system becomes formally equivalent to a
neutral and perfectly clean superconductor in a Zeeman field.

In this Letter we consider the effects of the long-range and
anisotropic interaction between atomic dipole moments in a Fermi
gas with population imbalance. The magnetic dipole interaction is
quite small for alkali atoms but can be considerably enhanced to
become experimentally observable for atoms with large dipole
moments \cite{Stuh05}. Another possibility is to control the
magnitude of electric dipole moments by applying external electric
field, as suggested in Ref. \cite{MY98}. In either case, we assume
that the dipole interaction represents a correction to the
dominant isotropic $s$-wave pairing. We will show that increasing
the population imbalance produces a novel nonuniform mixed-parity
(NMP) superfluid state, which has higher critical temperature than
the LOFF state.

We consider a homogeneous Fermi gas consisting of two different
species of atoms of equal mass $m$. For instance, in ${}^6$Li one
can have a mixture of the hyperfine states
$|+\rangle\equiv|m_s=1/2,m_i=1\rangle$ and
$|-\rangle\equiv|m_s=1/2,m_i=0\rangle$ \cite{SHSH96}, where $m_s$
and $m_i$ are the electron and nuclear spin projections
respectively. The free-particle Hamiltonian has the form
$H_0=\sum_{\bk}(\xi_{\bk}\delta_{\alpha\beta}-h\sigma_{3,\alpha\beta})
c^\dagger_{\bk\alpha}c_{\bk\beta}$, where
$\xi_{\bk}=(\bk^2-k_F^2)/2m$, $k_F$ is the Fermi momentum, $h$ is
the Fermi surface splitting due to unequal particle concentrations
in the states labelled by $\alpha,\beta=\pm$, and $\sigma_3$ is
the Pauli matrix (we set $\hbar=k_B=1$). We assume that the atomic
dipoles are fully polarized, e.g. by an external magnetic field
along the $z$-axis: $\bm{\mu}=\mu_d\hat z$. The interaction
between two atoms at distance $\bR$ is
\begin{equation}
\label{V-int}
    U(\bR)=-\lambda_0\delta(\bR)+\mu_d^2\frac{1-3\hat R_z^2}{R^3}.
\end{equation}
where the first term describes a short-range attraction with the
coupling constant $\lambda_0=4\pi|a|/m>0$, and the second term is
the dipole interaction. The $s$-wave scattering length $a<0$
depends on the field, diverging in the strong-coupling regime near
the Feshbach resonance. The effects of the dipole interaction
(\ref{V-int}) have been extensively studied for Bose gases, see
e.g. Ref. \cite{YY04} and the references therein.

In the BCS regime the pairing Hamiltonian, which takes into
account the possibility of the Cooper pairs having a nonzero
momentum $\bq$, can be written in the following form:
\begin{eqnarray}
\label{H-int}
    H_{int}&=&\frac{1}{2}\sum_{\bk,\bk',\bq}V(\bk,\bk')
    \Gamma_{\alpha\beta\gamma\delta}\nonumber\\
    &&\times c^\dagger_{\bk+\frac{\bq}{2},\alpha}c^\dagger_{-\bk+\frac{\bq}{2},\beta}
     c_{-\bk'+\frac{\bq}{2},\gamma}c_{\bk'+\frac{\bq}{2},\delta},
\end{eqnarray}
where $V(\bk,\bk')=-\lambda_0+4\pi
\mu_d^2(k_z-k'_z)^2/|\bk-\bk'|^2$ is the Fourier transform of Eq.
(\ref{V-int}) \cite{contact}. The matrix $\Gamma$ has the
following nonzero elements:
$\Gamma_{++++}=\Gamma_{+--+}=\Gamma_{-++-}= \Gamma_{----}=1$,
which is due to the fact that the pairing interaction depends only
on the total particle densities $n(\br)=n_+(\br)+n_-(\br)$.

Treating the Hamiltonian (\ref{H-int}) in the mean-field
approximation, one arrives at the self-consistency equation which
contains a divergent momentum integral due to the absence of
ultraviolet cutoff in $V(\bk,\bk')$. This divergence is removed by
the renormalization of the interaction potential \cite{GMB61}, in
which the bare interaction is replaced by the two-particle
scattering amplitude, and an effective cutoff $\omega_c$, of the
order of the Fermi energy $\epsilon_F$, is introduced. We use a
BCS-like sharp cutoff and represent the pairing interaction in a
factorized form:
$V(\bk,\bk')=-\theta(|\xi_{\bk}|-\omega_c)\theta(|\xi_{\bk'}|-\omega_c)
\sum_i\lambda_i\phi_i(\hat\bk)\phi^*_i(\hat\bk')$. The symmetry
factors $\phi_i(\hat\bk)$ are the eigenfunctions and the coupling
constants $-\lambda_i$ the eigenvalues of the operator $\hat V$
defined by the kernel $V(k_F\hat\bk,k_F\hat\bk')$. The symmetry
factors are even (odd) for the singlet (triplet) pairing and
satisfy the orthonormality condition:
$\langle\phi^*_i(\hat\bk)\phi_j(\hat\bk)\rangle=\delta_{ij}$,
where the angular brackets denote the average over the spherical
Fermi surface.

The effect of the dipole interaction on superfluidity is two-fold:
in addition to breaking the rotational symmetry of the system and
introducing gap anisotropy, it also leads to the possibility of
triplet pairing. We include one singlet and one triplet pairing
channel with the highest critical temperatures. One can show that
the lowest eigenvalues of $\hat V$ in both cases are achieved for
the eigenfunctions that do not depend on the azimuthal angle:
$\phi(\hat\bk)=f(\cos\theta)$, where the function $f(s)$ satisfies
the integral equation $\int_{-1}^1ds'K(s,s')f(s')=-\lambda f(s)$,
with the kernel $K(s,s')=-\lambda_0/2+\pi \mu_d^2|s-s'|$. In the
singlet case the solution is $f(s)\propto\cos(\kappa s)$, while in
the triplet case $f(s)\propto\sin(\kappa s)$, where
$\kappa=\sqrt{2\pi\mu_d^2/\lambda}$. For the singlet pairing, the
normalized symmetry factor has the form
\begin{equation}
\label{phi-even}
    \phi_s(\hat\bk)=\sqrt{\frac{2}{1+\frac{\sin 2\kappa_s}{2\kappa_s}}}
    \cos(\kappa_s\cos\theta),
\end{equation}
where $\kappa_s$ satisfies the equation
\begin{equation}
\label{kappa-eq}
    \kappa_s\tan\kappa_s=\frac{2\pi\mu_d^2}{\lambda_0-2\pi\mu_d^2},
\end{equation}
from which one obtains $\lambda_s=2\pi\mu_d^2/\kappa_s^2$. In the
limit of weak dipole interaction, $\mu_d^2\ll\lambda_0$, we find
$\phi_s(\hat\bk)\simeq 1+\kappa_s^2(1-3s^2)/6$, and
$\lambda_s\simeq\lambda_0-2\pi\mu_d^2$, i.e. the dipole
interaction leads to a downward renormalization of the coupling
constant in the singlet channel. For the triplet pairing, one can
show that the lowest eigenvalue corresponds to $\kappa=\pi/2$, so
that
\begin{equation}
\label{phi-odd}
    \phi_t(\hat\bk)=\sqrt{2}\sin\left(\frac{\pi}{2}\cos\theta\right),
\end{equation}
and $\lambda_t=(8/\pi)\mu_d^2$. Note that the coupling constant
and the symmetry factor in the triplet channel are not sensitive
to the value of $\lambda_0$ and therefore are the same as in the
pure dipolar case \cite{BMRS02}.

The singlet and triplet contributions to the pairing Hamiltonian
(\ref{H-int}) can now be explicitly separated:
\begin{equation}
\label{H-st}
    H_{int}=-\lambda_s\sum_{\bq}B_s^\dagger(\bq)B_s(\bq)
    -\lambda_t\sum_{\bq}\mathbf{B}_t^\dagger(\bq)\mathbf{B}_t(\bq),
\end{equation}
where
$B^\dagger_s(\bq)=(1/2)\sum_{\bk}\phi_s(\hat\bk)(i\sigma_2)_{\alpha\beta}
c^\dagger_{\bk+\bq/2,\alpha}c^\dagger_{-\bk+\bq/2,\beta}$ and
$\mathbf{B}^\dagger_t(\bq)=(1/2)\sum_{\bk}\phi_t(\hat\bk)(i\bm{\sigma}\sigma_2)_{\alpha\beta}
c^\dagger_{\bk+\bq/2,\alpha}c^\dagger_{-\bk+\bq/2,\beta}$ are the
pair creation operators. The representation (\ref{H-st})
emphasizes the full invariance of the pairing interaction with
respect to rotations in the ``pseudospin'' space spanned by the
states $|+\rangle$ and $|-\rangle$. Decoupling $H_{int}$ yields
the gap function
\begin{equation}
\label{Delta}
    \Delta_{\alpha\beta}(\hat\bk,\br)=(i\sigma_2)_{\alpha\beta}
    \psi(\br)\phi_s(\hat\bk)+(i\bm{\sigma}\sigma_2)_{\alpha\beta}
    \mathbf{d}(\br)\phi_t(\hat\bk),
\end{equation}
where the complex scalar $\psi$ and the complex vector
$\mathbf{d}$ are the order parameters of the singlet and the
triplet pairing respectively.

In the absence of population imbalance $h=0$, and the critical
temperatures for the singlet and triplet pairing are given by the
standard BCS expressions: $T_a\equiv
T^{(a)}_{c0}=(2e^\mathbb{C}/\pi)\omega_ce^{-1/N_F\lambda_a}$,
where $a=s,t$ and $N_F$ is the density of states at the Fermi
surface per one hyperfine state (all three components of
$\mathbf{d}$ have the same critical temperature). Of the most
interest to us is the limit of weak dipole interaction, in which
$T_s\gg T_t$.

The phase diagram at $h\neq 0$ can be obtained from the free
energy, which is represented as an expansion in powers of the
order parameter components: ${\cal F}={\cal F}_0+{\cal
F}_2[\psi,\mathbf{d}]+...$ (${\cal F}_0$ is the free energy in the
normal state). To find the critical temperature $T_c(h)$ of the
second-order phase transition into the superfluid state, or
inversely the critical Fermi-surface splitting $h_c(T)$, it is
sufficient to keep only the quadratic terms in ${\cal F}$. Using
Eq. (\ref{H-st}) we find that the contributions to ${\cal F}_2$
from $\psi$ and $d_z$, which describe the pairing between fermions
of different species, are decoupled from $d_\pm=(d_x\pm
id_y)/\sqrt{2}$, which correspond to the intra-species pairing:
\begin{eqnarray}
\label{F2}
    {\cal F}_2=\sum_\bq
    \biggl[\left(\begin{array}{cc}
    \psi^*_\bq & d^*_{z,\bq} \\
    \end{array}\right)
    \left(\begin{array}{cc}
    A_{ss} & A_{st} \\
    A_{ts} & A_{tt} \\
    \end{array}\right)
    \left(\begin{array}{c}
    \psi_\bq \\
    d_{z,\bq} \\
    \end{array}\right)\nonumber\\
    +A_+|d_{+,\bq}|^2+A_-|d_{-,\bq}|^2\biggr].
\end{eqnarray}
The coefficients in this expression have the following form:
$A_{ab}=A_{ba}=N_F[\delta_{ab}\ln(T/T_a)+I_{ab}]$,
$A_\pm=N_F[\ln(T/T_t)+I_{tt}(h=0)\pm\delta I]$,
\begin{eqnarray}
\label{Iab}
    I_{ab}=
    \biggl\langle\phi_a(\hat\bk)\phi_b(\hat\bk)\re\Psi\biggl(\frac{1}{2}+
    \frac{iW}{4\pi T}\biggr)\biggr\rangle
    -\delta_{ab}\Psi\left(\frac{1}{2}\right),
\end{eqnarray}
where $\phi_{s,t}(\bk)$ are defined by Eqs. (\ref{phi-even}) and
(\ref{phi-odd}), $\Psi(x)$ is the digamma function,
$W=v_F\hat\bk\bq-2h$, $v_F$ is the Fermi velocity, and $\delta
I\propto N^\prime_Fh$ is proportional to the difference between
the densities of states at the Fermi levels for the two fermionic
species. Since at $h\neq 0$ the symmetry in the ``pseudospin''
space is reduced to rotations about the $z$-axis, different
components of $\mathbf{d}$ appear at different temperatures.

The critical temperatures for $d_+$ and $d_-$ are found from the
equations $A_+=0$ and $A_-=0$ respectively. If one neglects the
band asymmetry $N_F'$ then the phase transition is not affected by
the population imbalance, otherwise $T_{c,\pm}(h)=T_t\mp O(\delta
I)$, so that a nonunitary triplet state with $d_+=0,d_-\neq 0$ is
realized.

The effect of the Fermi-surface splitting on the other two
components of the order parameter, $\psi$ and $d_z$, is more
interesting. Because of the presence of the off-diagonal matrix
elements in $\hat A$, see Eq. (\ref{Iab}), the singlet and triplet
pairing channels can be mixed at $\bq\neq 0$ and $h\neq 0$,
producing the NMP state, in which both $\psi$ and $d_z$ are
nonzero. The critical temperature is obtained from the equation
$\det\hat A(\bq)=0$, after maximization with respect to $\bq$. The
calculation is facilitated by the observation that the functions
(\ref{Iab}) can be expressed in terms of two dimensionless
variables $\bQ=v_F\bq/2h$ and $z=h/2\pi T$. At any given $0\leq
z\leq\infty$, we find
\begin{equation}
    T_c(h)=T_s\max\limits_\bQ e^{{\cal I}(\bQ,z)},
\end{equation}
where
\begin{eqnarray*}
    {\cal I}=-\frac{1}{2}(I_{ss}+I_{tt}+r)+\frac{1}{2}\sqrt{(I_{ss}-I_{tt}-r)^2+4I_{st}^2}.
\end{eqnarray*}
The critical band splitting is given by $h_c(T)=2\pi zT_c$. If the
maximum of ${\cal I}$ is achieved at $\bQ=\bQ_c$, then the wave
vector of the superfluid instability is $\bq_c=2h_c\bQ_c/v_F$. The
parameter $r=\ln(T_s/T_t)>0$ characterizes the relative strength
of pairing in the singlet and triplet channels.

In the absence of dipole interaction, $r=\infty$, the pairing is
isotropic, and ${\cal
I}(\bQ,z)=\Psi(1/2)-\langle\re\Psi(1/2-iz+iz\hat\bk\bQ)\rangle$.
At $z\leq z_{\mathrm{LOFF}}\simeq 0.30$, this function has a
maximum at $\bQ=0$, therefore the phase transition occurs into the
uniform superfluid state. At $z>z_{\mathrm{LOFF}}$, i.e. at
$T<T_{\mathrm{LOFF}}\simeq 0.56T_s$, the maximum of ${\cal I}$ is
achieved at $|\bQ|\neq 0$, which corresponds to the LOFF state.

Since the dipole interaction lifts the spherical degeneracy of the
critical temperature, the cases $\bq\perp\hat z$ and
$\bq\parallel\hat z$ should be considered separately. In our
numerical analysis we use the following values of the parameters,
appropriate for the weak dipole interaction in the BCS limit:
$(N_F\lambda_0)^{-1}=1.0$ and $\mu_d^2/\lambda_0=0.1$, which gives
$T_t\simeq 0.02T_s$ and $r\simeq 3.93$.

For $\bq=q\hat z$, we find that at $z\leq z_{\mathrm{NMP}}\simeq
0.23$ the maximum of ${\cal I}$ is at $\bQ=0$, therefore
$I_{st}=0$ and the phase transition occurs into the uniform
singlet state. In contrast, at $z>z_{\mathrm{NMP}}$, i.e. at
$T<T_{\mathrm{NMP}}\simeq 0.68T_s$, the critical temperature has
two degenerate maxima at $\bQ=\pm Q_c\hat z\neq 0$, which
corresponds to the phase transition into the state
\begin{equation}
\label{NMP}
    \left(\begin{array}{c}
    \psi \\
    d_z \\
    \end{array}\right)=
    \eta_1
    \left(\begin{array}{c}
    1 \\
    \rho \\
    \end{array}\right)e^{iq_cz}+
    \eta_2
    \left(\begin{array}{c}
    1 \\
    -\rho \\
    \end{array}\right)e^{-iq_cz}.
\end{equation}
Here $\rho=-I_{st}/[\ln(T_c/T_t)+I_{tt}]$, and the weights
$\eta_{1,2}$ of the two plane-wave components are determined by
minimizing the full nonlinear free energy ${\cal F}$, see below.
For $\bq\perp\hat z$, there is no singlet-triplet mixing terms,
and at $T<T_{\mathrm{LOFF}}$ one obtains the LOFF state with
$\psi$ modulated in the equatorial plane and $d_z=0$. The
transition line for this state is determined by the maximum of
${\cal I}(\bQ,z)=-I_{ss}$, which yields the critical temperature
slightly higher than in the isotropic LOFF case but still lower
than in the NMP case.

\begin{figure}
    \includegraphics[width=7.2cm]{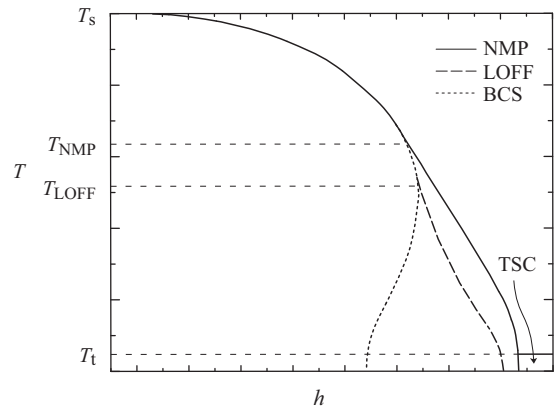}
    \caption{Critical temperature vs Fermi-surface splitting
    in a homogeneous Fermi gas with dipole interaction.
    The solid line corresponds to the transition into
    the NMP state with $\bq\parallel\hat z$, the dashed line corresponds to the
    LOFF state with $\bq\perp\hat z$, and the short-dashed line is the instability line
    of the uniform BCS singlet state. At the lowest temperatures, $T<T_t$, the nonunitary
    triplet state (TSC) is realized.}
    \label{Phase Diagram}
\end{figure}

Thus we come to the conclusion that it is the NMP state
(\ref{NMP}) that has the highest critical temperature below
$T_{\mathrm{NMP}}$, see Fig. \ref{Phase Diagram}. One can show
that as $r$ varies from $+\infty$ (no dipole interaction, $T_t=0$)
to $0$ (strong dipole interaction, $T_t=T_s$), $T_{\mathrm{NMP}}$
moves from $T_{\mathrm{LOFF}}$ towards $T_s$. The parameter $\rho$
is the measure of the triplet component admixture, which is zero
at $T=T_{\mathrm{NMP}}$ and increases as temperature decreases
[for the parameter values used above, $\rho(T=0)\simeq 0.23$].

In order to determine the spatial structure of the NMP phase just
below the critical temperature, one needs to evaluate the
fourth-order terms in the free energy density:
\begin{equation}
\label{F4}
    F_4=\beta_1(|\eta_1|^4+|\eta_2|^4)+\beta_2|\eta_1|^2|\eta_2|^2.
\end{equation}
The coefficients $\beta_{1,2}$ are functions of temperature. The
above expression is positive definite if $\beta_1>0$,
$\beta_2>-2\beta_1$. At $\beta_2>0$ the minimum is achieved in the
state $(\eta_1,\eta_2)\sim(1,0)$ or $(0,1)$, while at $\beta_2<0$
one has $(\eta_1,\eta_2)\sim(1,e^{i\varphi})$, where $\varphi$ is
an arbitrary phase. Near $T_{\mathrm{NMP}}$ one can set
$\rho=q_c=0$ in the expressions for $\beta_{1,2}$, which gives
$\beta_1=\beta_2/4=-(N_F\langle\phi_s^4\rangle/32\pi^2T_{\mathrm{NMP}}^2)\re
\Psi^{\prime\prime}(1/2-iz_{\mathrm{NMP}})>0$. Therefore, the NMP
phase transition in the vicinity of $T_{\mathrm{NMP}}$ is of the
second order, with $(\psi,d_z)\sim(1,\rho)e^{iq_cz}$. The
determination of the full phase diagram, including the spatial
structure of the order parameter at all $T<T_{\mathrm{NMP}}$, is
much more difficult, because of the competition between the NMP,
LOFF, and uniform BCS phases. We leave this problem for future
studies.

The origin of the NMP instability can be elucidated using the
Ginzburg-Landau expansion of the free energy (\ref{F2}) at small
$h$ and $\bq=q\hat z$ in the vicinity of $T_s$. In addition to the
usual uniform and gradient terms, the energy density contains
mixed-parity terms, which are linear in gradients \cite{helical}:
\begin{equation}
\label{GL energy}
    F_2=\psi^*\hat A_s\psi+d_z^*\hat A_td_z+i\tilde
    Kh(\psi^*\nabla_zd_z+d_z^*\nabla_z\psi),
\end{equation}
where $\hat A_a=N_F(T-T_a)/T_a-K_a\nabla^2_z$,
$K_a=7\zeta(3)N_F\langle\phi_a^2 v_z^2\rangle/16\pi^2T_a^2$, and
$\tilde K=7\zeta(3)N_F\langle\phi_s\phi_t v_z\rangle/4\pi^2T_s^2$.
Because of the last term, which favors a spatial modulation of the
order parameter, at
$h>h_{\mathrm{NMP}}=\sqrt{N_FK_s(T_s-T_t)/T_t\tilde K^2}$ the
superfluid phase transition is of the NMP type. In contrast, the
onset of the singlet LOFF instability is marked by the coefficient
$K_s$ changing sign at $h_{\mathrm{LOFF}}>h_{\mathrm{NMP}}$.

To summarize, we found that in a homogeneous Fermi gas with
$s$-wave attraction, the atomic dipole interaction produces
triplet pairs that can mix with the singlet order parameter in a
spatially modulated superfluid state. If the densities of atoms in
different hyperfine states are unequal then the nonuniform
mixed-parity state has higher critical temperature than the LOFF
state, even at weak dipole interaction. To relate these findings
with the cold atom experiments the effects of the trapping
potential should be included. Another interesting open question
concerns the fate of the NMP state in the strong-coupling regime
near the BCS-BEC crossover.

Finally, we would like to remark that the possibility of a
nonuniform singlet-triplet mixing is not restricted to polarized
Fermi gases. For example, in a fully isotropic system the pairing
interaction $V(\bk,\bk')$ in Eq. (\ref{H-int}) depends only on the
angle between $\bk$ and $\bk'$ and can therefore be factorized in
terms of the spherical harmonics $Y_{lm}(\hat\bk)$. In the singlet
$s$-wave channel the order parameter is a scalar $\psi$, with
$\phi_s(\hat\bk)=Y_{00}(\hat\bk)=1$, while in the triplet $p$-wave
channel the order parameter is now represented by three vectors
$\mathbf{d}_m$, with $m=0,\pm 1$, corresponding to
$\phi_{t,m}(\hat\bk)=Y_{1m}(\hat\bk)$. When expressed in terms of
the Cartesian components of $\hat\bk$, the triplet order parameter
becomes a 3$\times$3 complex matrix $A_{i\alpha}$ \cite{VW90}. In
the presence of magnetic field $\mathbf{h}$, there is a
contribution to the free energy density of the form
$ih_i(\psi^*\nabla_\alpha A_{i\alpha}-\mathrm{c.c.})$, which is
invariant under all required symmetry operations: orbital and spin
rotations, time reversal, and inversion.  Similarly to the last
term in Eq. (\ref{GL energy}) this can lead to the NMP instability
at sufficiently high $h$.

This work was supported by the Natural Sciences and Engineering
Research Council of Canada.


\begin{thebibliography}{99}
\bibitem{contact}
The Fourier transform of the dipole interaction also contains a
constant term, which has been included in the renormalization of
$\lambda_0$.
\end{thebibliography}
\end{document}